\documentstyle[titlepage,12pt]{article}
\newcommand{\beq}{\begin{equation}}
\newcommand{\eeq}{\end{equation}}

\newcommand{\eq}[1]{eq.(\ref{#1})}
\title {
\begin{flushright} PNPI-1942\\
hep-ph/9312337\\
December 1993
\end{flushright}
\bigskip\bigskip\bigskip
CORRECTIONS TO HYPERFINE SPLITTING AND LAMB SHIFT
INDUCED BY THE OVERLAPPING TWO-LOOP ELECTRON
SELF-ENERGY INSERTION IN THE ELECTRON LINE}

\author {Michael I.  Eides \\
Petersburg Nuclear Physics Institute,\\
Gatchina, St.Petersburg 188350, Russia\thanks{E-mail address:
eides@lnpi.spb.su}\\
and
\medskip\\
\and Savely G. Karshenboim \hspace{.5cm} and \hspace{.5cm} Valery A.
Shelyuto\\
D. I.  Mendeleev Institute of Metrology, \\
St.Petersburg 198005, Russia}
\date{}

\begin{document}
\maketitle
\tableofcontents
\begin{abstract}

Contributions to HFS and to the Lamb shift intervals of order
$\alpha^2(Z\alpha)^5$ induced by the graph with the two-loop overlapping
electron self-energy diagram inserted in the electron line are considered.
Explicit expression for the overlapping two-loop self-energy diagram in
the Fried-Yennie gauge is obtained. Contributions both to HFS and Lamb
shift induced by the diagram containing such subgraph are calculated.
\end{abstract}

\section{Introduction}

Calculation of contributions of order $\alpha^2(Z\alpha)^5m$ to hyperfine
splitting (HFS) and Lamb shift induced by the two-loop radiative
photon insertions in the electron line was initiated in the previous paper
\cite{ekse1}. We have calculated there the contributions induced by all
graphs containing one-loop electron self-energy diagram as a subgraph, by
the graph containing two one-loop vertices, and also the contribution
induced by the rainbow two-loop electron self-energy insertion in the
electron line\footnote{We use a chance to correct an error made in
\cite{ekse1} in the calculation of the rainbow diagram contribution to the
Lamb shift. Due to a normalization error the result cited in \cite{ekse1} is
two times larger than the correct one $$\Delta E_{L}^{h}
=\frac{153}{80}\frac{\alpha^2(Z\alpha)^5}{\pi n^3}
\left(\frac{m_r}{m}\right)^3\:m\:.$$ We are deeply grateful to Dr.
Pachucki who attracted our attention to this error.}.  Calculation of the
contribution induced by the overlapping two-loop electron self-energy
insertion (see Fig.1) is presented below. Consideration of this contribution
is impeded by all the usual difficulties which are connected with the
presence of overlapping infinities in this case and the necessity to perform
all necessary subtractions. In Section \ref{total} we obtain explicit
expressions for the renormalized overlapping diagram contribution to the
self-energy operator in the Fried-Yennie (FY) gauge in the form of the
Feynman parameter integral. These expressions are used later for calculation
of the contributions to HFS and Lamb shift. It should be mentioned
that taking into account also expressions for the rainbow and one-particle
reducible \cite{ekse1} diagrams we have explicit general expression for the
two-loop mass-operator in the FY gauge. We are aware of only one other
calculation of the two-loop mass operator existing in the literature
\cite{sabry} which was performed in the Feynman gauge. We hope that the
explicit expression in the FY gauge presented below may find applications
also in other problems, e.g.  in calculation of the positronium decay rate
\cite{adkins}.

\section{General Expression for the Overlapping Self-Energy Operator}

Consider contribution to the two-loop self-energy operator induced by the
diagram with overlapping photons. First of all one has to perform
subtractions of potential infinities in internal vertices. Explicit
expression for such internally subtracted contribution to the mass-operator
in the FY gauge has the form

\beq          \label{sov}
\Sigma(p)=(\frac{\alpha}{4\pi})^2\int\frac{d^4l}{i\pi^2}
\int\frac{d^4q}{i\pi^2}\frac{1}{l^2q^2}(g^{\alpha\lambda}+2\frac{l^\alpha
l^\lambda}{l^2})(g^{\beta\sigma}+2\frac{q^\beta q^\sigma}{q^2})
\eeq
\[
\gamma_\alpha\{\frac{\hat p+\hat l+m}{D(p+l)}\gamma_\beta\frac{\hat p+\hat
l+\hat q+m}{D(p+l+q)}\gamma_\lambda\frac{\hat p+\hat q+m}{D(p+q)}
\]
\[
-\frac{\hat p_0+\hat l+m}{D(p_0+l)}\gamma_\beta\frac{\hat p_0+\hat
l+m}{D(p_0+l)}\gamma_\lambda\frac{\hat p+\hat q+m}{D(p+q)}
\]
\[
-\frac{\hat p+\hat l+m}{D(p+l)}\gamma_\beta\frac{\hat p_0+\hat
q+m}{D(p_0+q)}\gamma_\lambda\frac{\hat p_0+\hat q+m}{D(p_0+q)}\}
\gamma_\sigma
\]

where

\[
D(p)=p^2-m^2
\]

and $\hat p_0=m$ is the mass-shell momentum in subtracted terms.

There is a certain subtlety connected with this mass-shell limit. Although
renormalization procedure in the FY gauge is performed below without
introduction of the infrared photon mass (see, for more details \cite{ann1})
there are spurious infrared divergences on the intermediate stages of
calculations on the mass-shell which may produce finite but
discontinuous on the mass shell results if the mass-shell limit is
taken in the naive way prior to calculation of the integrals.  To avoid
these problems one has to perform calculation even of the subtracted terms
with slightly off mass-shell external momentum (in this case with off
mass-shell momentum $p_0$) and only after calculation of all
infrared unsafe integrals to take the mass-shell limit. In this way one
preserves continuity of all physical results on the mass shell and obtains
correct results.

We would like to obtain the expression for the overlapping diagram
contribution to the self-energy operator in the form of the integral
representation with minimal number of the Feynman parameters. To reduce the
number of the parameters we transform the integrand in \eq{sov} with the
help of trivial identity

\beq
\hat p+\hat l+\hat q+m=(\frac{\hat p+m}{2}+\hat l)+(\frac{\hat p+m}{2}+\hat
q)
\eeq

and obtain

\beq           \label{trick}
\Sigma(p)=(\frac{\alpha}{4\pi})^2\int\frac{d^4l}{i\pi^2}
\int\frac{d^4q}{i\pi^2}\frac{1}{l^2q^2}(g^{\alpha\lambda}+2\frac{l^\alpha
l^\lambda}{l^2})(g^{\beta\sigma}+2\frac{q^\beta q^\sigma}{q^2})
\eeq
\[
\frac{1}{2}\{\gamma_\sigma\frac{\hat p+\hat
q+m}{D(p+q)}\gamma_\lambda {\cal R}_\beta\gamma_\alpha
+\gamma_\alpha {\cal R}_\beta\gamma_\lambda\frac{\hat p+\hat
q+m}{D(p+q)}\gamma_\sigma\},
\]

where

\beq
{\cal R}_\beta=\frac{(\hat p+\hat l+m)\gamma_\beta(\hat p+\hat
2l+m)}{D(p+l)D(p+l+q)}-2\frac{(\hat p_0+\hat l+m)
\gamma_\beta(\hat p_0+\hat l+m)}{D^2(p_0+l)}.
\eeq

One may check that both terms in the braces in \eq{trick} produce
after integration coinciding results, so we consider below only integral
of the second term and simply double the result. Initial expression for
further transformations has the form

\beq      \label{rearrs}
\Sigma(p)=(\frac{\alpha}{4\pi})^2\int\frac{d^4q}{i\pi^2}\frac{1}{q^2}
(g^{\beta\sigma}+2\frac{q^\beta
q^\sigma}{q^2})R_\beta(p,q)\frac{\hat p+\hat q+m}{D(p+q)}\gamma_\sigma,
\eeq

where

\beq     \label{intern}
R_\beta(p,q)=\int\frac{d^4l}{i\pi^2}\frac{1}{l^2}(g^{\alpha\lambda}
+2\frac{l^\alpha l^\lambda}{l^2})\gamma_\alpha {\cal R}_\beta\gamma_\lambda.
\eeq

Integration over $l$ in \eq{intern} is convergent but separate terms in the
integrand produce  logarithmically divergent results. We slightly rearrange
integrand in order to separate compensating divergences

\beq          \label{repr}
R_\beta(p,q)=\int\frac{d^4l}{i\pi^2}\{\frac{1}{l^2D(p+l)D(p+l+q)}
\eeq
\[
[(g^{\alpha\lambda}+2\frac{l^\alpha l^\lambda}{l^2})\gamma_\alpha
(\hat p+\hat l+m)\gamma_\beta(\hat p+2\hat
l+m)\gamma_\lambda-6l^2\gamma_\beta]
\]
\[
-\frac{2}{l^2D^2(p_0+l)}[(g^{\alpha\lambda}+2\frac{l^\alpha l^\lambda}{l^2})
\gamma_\alpha (\hat p_0+\hat l+m)\gamma_\beta(\hat p_0+\hat l+m)
\gamma_\lambda-3l^2\gamma_\beta]
\]
\[
+6\gamma_\beta[\frac{1}{D(p+l)D(p+l+q)}-\frac{1}{D^2(p_0+l)}]\}
\]
\[
\equiv R^{(1)}_\beta+R^{(2)}_\beta+R^{(3)}_\beta.
\]

Representation in eq.(\ref{repr}) is very convenient for further
transformations as each of the terms $R^{(i)}_\beta$ (after substitution
in \eq{rearrs}) does not need more than four integration parameters to
represent it in the form of the integral and each of these terms is a
convergent integral in respect with integration over $l$ which will be the
first integral to perform. We will consider transformations of each of the
terms in \eq{repr} separately.

\section {Calculation of the Contribution to the Mass Operator Induced
by the Term $R^{(1)}_\beta$}

Let us consider first contribution to the mass operator induced by term
$R^{(1)}_\beta$ in eq.(\ref{repr}). After ordering over powers of the
integration momentum $l$ we obtain

\beq           \label{1integr}
R^{(1)}_\beta=2\int\frac{d^4l}{i\pi^2}\frac{1}{l^2D(p+l)D(p+q+l)}
\{2p_\beta(3m-\hat p+\frac{\hat l\hat p\hat l}{l^2})
+(p^2-m^2)(\gamma_\beta-\frac{\hat l\gamma_\beta\hat
l}{l^2})
\eeq
\[
+[6ml_\beta-2\hat l\gamma_\beta\hat p-\hat
p\gamma_\beta\hat l+2\hat l(\hat p+m)\gamma_\beta+\gamma_\beta(\hat
p+m)\hat l]-[2\hat l\gamma_\beta\hat l+l^2\gamma_\beta]\}.
\]

Next we combine denominators with the help of the identity

\beq       \label{combg}
(1-x)l^2+x[(1-z)D(p+l)+zD(p+l+q)]=(l+xQ)^2-x\Delta,
\eeq

where

\[
Q=p+qz,
\]
\[
\Delta=m^2-p^2(1-x)-q^2z(1-xz)-2pqz(1-x).
\]

After shift of the integration variable $l\rightarrow l-xQ$ we perform
momentum integration  which leads to the expression

\beq       \label{bmass}
R^{(1)}_\beta=2\int_0^1dx\int_0^1dz
\{2p_\beta[\frac{\hat p(2-x)-3m}{\Delta}+x(1-x)\frac{\hat Q\hat p\hat
Q}{\Delta^2}]
\eeq
\[
-(p^2-m^2)[(2-x)\frac{\gamma_\beta}{\Delta}+x(1-x)\frac{\hat
Q\gamma_\beta\hat Q}{\Delta^2}]+[\frac{x}{\Delta}C^{(1)}_\beta
+\frac{x^2}{\Delta}C^{(2)}_\beta]\}.
\]
\[
\equiv R^{(1a)}_\beta+R^{(1b)}_\beta+R^{(1c)}_\beta.
\]

where

\beq
C^{(1)}_\beta=6mQ_\beta-2\hat Q\gamma_\beta\hat p-\hat p\gamma_\beta\hat Q
+2\hat Q(\hat p+m)\gamma_\beta+\gamma_\beta(\hat p+m)\hat Q,
\eeq
\[
C^{(2)}_\beta=2\hat Q\gamma_\beta\hat Q+Q^2\gamma_\beta.
\]

Each power of momentum $l$ in the numerator of the integrand in
\eq{1integr} produces respective power of factor $x$ in the numerator of the
integrand in \eq{bmass}. We will consider calculation of the terms in
different brackets on the right hand side (i.e. terms $R^{(1i)}_\beta$)
separately.

\subsection{Contribution to the Self-Energy Operator Induced by the Term
$R^{(1a)}_\beta$}

It is easy to see that the expression in \eq{bmass} leads to the
contribution to the unrenormalized self-energy operator which remains finite
even on the mass shell. However, we would like to obtain expression for the
renormalized (doubly subtracted) self-energy operator and each subtraction
inserts (after integration over $q$) additional power of $x$ in the
denominator, making integration over $x$ potentially unsafe. Only term
$R^{(1a)}_\beta$ in \eq{1integr} seems to produce infrared divergency on the
mass shell after subtraction. To get rid of this apparent infrared
divergency we first separate infrared unsafe terms in the expression for
this term

\beq
R^{(1a)}_\beta=4p_\beta\int_0^1dx\int_0^1dz\{[\frac{\hat
p(2-x)-3m}{\Delta}+x(1-x)\frac{\hat pQ^2}{\Delta^2}]
\eeq
\[
+x(1-x)\frac{\hat Q\hat p\hat Q-\hat pQ^2}{\Delta^2}\}.
\]

Next we use trivial identity which is the result of integration by parts and
the relation $\partial \Delta/\partial x=Q^2$

\beq          \label{intpart}
\int_0^1dxx(1-x)\frac{Q^2}{\Delta^2}=\int_0^1dx\frac{1-2x}{\Delta}
\eeq

and obtain

\beq                \label{bb1}
R^{(1a)}_\beta=4p_\beta\int_0^1dx\int_0^1dz\{3\frac{\hat p(1-x)-m}{\Delta}
+x(1-x)\frac{\hat Q\hat p\hat Q-\hat pQ^2}{\Delta^2}\}.
\eeq

This new representation is completely infrared safe and admits subtraction
on the mass-shell. Respective contribution to the mass operator has the form

\beq        \label{mass1}
\Sigma^{(1a)}(p)=4(\frac{\alpha}{4\pi})^2
\int_0^1dx\int_0^1dz\int\frac{d^4q}{i\pi^2}\{3\frac{\hat p(1-x)-m}{\Delta}
\eeq
\[
+x(1-x)\frac{\hat Q\hat p\hat Q-\hat
pQ^2}{\Delta^2}\}\frac{\hat p+\hat q+m}{q^2D(p+q)}(2\hat p+\frac{\hat q\hat
p\hat q}{q^2}).
\]

Next we introduce Feynman parameters with the help of the identity

\beq       \label{comb}
(1-t)q^2+t[u(-\frac{\Delta}{z(1-xz)})+(1-u)D(p+q)]=(q+p\eta t)^2-t\Omega,
\eeq

where

\beq                   \label{auxil}
\Omega=p^2\eta^2t+(m^2-p^2)(1-u)+\frac{m^2-p^2(1-x)}{z(1-xz)}u\equiv
\frac{m^2(\omega+\rho\xi)}{z(1-xz)},
\eeq
\[
\eta=1-\frac{x(1-z)u}{1-xz},
\]
\[
\rho=\frac{m^2-p^2}{m^2},
\]
\[
\xi=(1-x)u+z(1-xz)(1-u-\eta^2t),
\]
\[
\omega=ux+z(1-xz)\eta^2t.
\]

Denominators in \eq{mass1} are combined as follows

\beq                    \label{genr}
\frac{1}{\Delta^kD(p+q)q^{2n}}=\int_0^1du\int_0^1dt\frac{g(k,1,n)}{[(q+p\eta
t)^2-t\Omega]^{k+1+n}}\equiv G(k,1,n),
\eeq

where

\beq
g(1,1,0)=-\frac{1}{z(1-xz)},\: g(1,1,1)=-\frac{2t}{z(1-xz)},\:
g(1,1,2)=-\frac{6t(1-t)}{z(1-xz)},
\eeq
\[
g(2,1,1)=\frac{6ut^2}{z^2(1-xz)^2},\quad
g(2,1,2)=\frac{24ut^2(1-t)}{z^2(1-xz)^2}.
\]

We displayed in \eq{auxil} and \eq{genr} slightly more general formulae than
those which are necessary to perform momentum integration in \eq{mass1}.
These auxiliary equations will be extensively used below for momentum
integrations of other entries in \eq{rearrs}.

Contribution to the mass operator in \eq{mass1} has a rather simple form in
new notation

\beq                       \label{bb1mom}
\Sigma^{(1a)}(p)=(\frac{\alpha}{4\pi})^2\int_0^1dx\int_0^1dz
\int\frac{d^4q}{i\pi^2}\sum_{k,n=1,2}N_a(k,1,n)G(k,1,n),
\eeq

where

\beq
N_a(1,1,1)=12[\hat p(1-x)-m][2\hat p(\hat p+m)+(2\hat q\hat p+\hat p\hat
q)],
\eeq
\[
N_a(1,1,2)=12[\hat p(1-x)-m](\hat p+m)\hat q\hat p\hat q,
\]
\[
N_a(2,1,1)=4x(1-x)(\hat Q\hat p\hat Q-\hat pQ^2)[2\hat p(\hat p+m)+(2\hat
q\hat p+\hat p\hat q)],
\]
\[
N_a(2,1,2)=4x(1-x)(\hat Q\hat p\hat Q-\hat pQ^2)(\hat p+m)\hat q\hat p\hat
q.
\]

We want to obtain after momentum integration as small powers of factor
$\Omega$ in the denominator as possible since in this case next
integrations are more convergent and subtraction procedure is more
accessible. Due to this reason we separated in the numerators characteristic
structure $\hat Q\hat p\hat Q-\hat pQ^2$ which does not contain cubic in
external momentum $p$ terms after shift of integration momentum and leads
thus to smaller power of $\Omega$ after integration.

Next we shift integration momentum $q\rightarrow q-p\eta
t$ and obtain after straightforward integration in \eq{mass1}

\beq              \label{mss}
\Sigma^{(1a)}(p)=(\frac{\alpha}{4\pi})^2\int_0^1dx\int_0^1dz
\int_0^1du\int_0^1dt\frac{12}{z(1-xz)}\{[\hat p(1-x)-m]
\eeq
\[
\{\frac{1}{\Omega}[2\hat p(\hat p+m)-3p^2\eta t]-\hat p(\hat
p+m)[\frac{1-t}{\Omega}+t(1-t)\frac{p^2\eta^2}{\Omega^2}]\}
\]
\[
+\frac{x(1-x)u}{1-xz}\{-\frac{p^2t}{\Omega}[\hat p(1+2z\eta t)-2(\hat
p+m)z]-\frac{p^2zt(1-t)}{\Omega}(\hat p+3m)
\]
\[
+\frac{p^4(1-t)}{\Omega^2}[2(\hat
p-m)\eta t-(3\hat p-m)z\eta^2t^2]\}\}.
\]

One may get rid of denominator $\Omega^2$ with the help of integration by
parts over $t$, taking into account that $\partial\Omega/\partial
t=p^2\eta^2$ (compare \eq{intpart}). Integration by parts leads to
substitutions in \eq{mss}

\beq
\frac{t(1-t)p^2\eta^2}{\Omega^2}\rightarrow \frac{1-2t}{\Omega},
\eeq
\[
\frac{t^2(1-t)p^2\eta^2}{\Omega^2}\rightarrow \frac{t(2-3t)}{\Omega}.
\]

Note that although the right hand side in \eq{mss} is finite even on the
mass shell, separate terms in the integrand may lead to infrared divergent
contributions after subtraction. Integration by parts gave us the chance
to get rid of these would be IR divergence which could become dangerous for
the accessibility of the subtraction procedure.

Now expression in eq.(\ref{mss}) acquires the form

\beq                  \label{mass1a}
\Sigma^{(1a)}(p)=(\frac{\alpha}{4\pi})^2\int_0^1dx\int_0^1dz
\int_0^1du\int_0^1dt\frac{12}{m^2(\omega+\rho\xi)}\{3\hat pt[\hat p(1-x)-m]
\eeq
\[
[\hat p(1-\eta)+m]+\frac{x(1-x)u}{1-xz}p^2[2(\hat p-m)\frac{1-2t}{\eta}
-\hat pt(1+2z\eta t)+zt[m-5\hat p(1-2t)]]\}.
\]

This form is very suitable for subtraction which we postpone until all
other contributions to the mass operator would be obtained.

\subsection {Calculation of the Contribution to the Mass Operator Induced
by the Term $R^{(1b)}_\beta$}

Contribution to the self-energy operator corresponding to $R^{(1b)}_\beta$
has the form

\beq
\Sigma^{(1b)}(p)=-2(\frac{\alpha}{4\pi})^2(p^2-m^2)\int_0^1dx\int_0^1dz
\int\frac{d^4q}{i\pi^2}\{(2-x)\frac{\gamma_\beta}{\Delta}
\eeq
\[
+x(1-x)\frac{\hat
Q\gamma_\beta\hat Q}{\Delta^2}\}\frac{\hat p+\hat
q+m}{q^2D(p+q)}(\gamma^\beta+\frac{2\hat qq^\beta}{q^2}).
\]

Again as in the previous section we try to transform numerator of the
integrand to the form containing minimal power of external momentum $p$.
This goal is easily reached with the help of trivial identities

\beq
\hat Q\gamma_\beta\hat Q(\hat p+\hat q+m)\gamma^\beta=2Q^2(2\hat p+2\hat
q-m)+2(1-z)[\hat Q\hat q\hat Q-\hat qQ^2],
\eeq
\[
\hat Q\hat q\hat Q =\hat qQ^2+[\hat Q\hat q\hat Q-\hat qQ^2].
\]

Note that the term $\hat Q\hat q\hat Q-\hat qQ^2$ as well as the similar
term $\hat Q\hat p\hat Q-\hat pQ^2$ in the previous section is free of
cubic in the external momentum terms after shift of integration variable.
We get rid of the terms containing $Q^2$ in the numerator with the help of
integration by parts displayed in \eq{intpart}. We then obtain

\beq
\Sigma^{(1b)}(p)=-2(\frac{\alpha}{4\pi})^2(p^2-m^2)\int_0^1dx\int_0^1dz
\int\frac{d^4q}{i\pi^2}\sum_{k,n=1,2}N_b(k,1,n)G(k,1,n),
\eeq

where

\beq
N_b(1,1,1)=6[\hat q(1-2x)-\hat px+m(2-x)],
\eeq
\[
N_b(1,1,2)=6(1-x)\hat q\hat p\hat q,
\]
\[
N_b(2,1,1)=2x(1-x)(2-z)(\hat Q\hat q\hat Q-\hat qQ^2),
\]
\[
N_b(2,1,2)=2x(1-x)(\hat Q\hat q\hat Q-\hat qQ^2)(\hat p+m)\hat q.
\]

Performing next straightforward momentum integration one obtains

\beq              \label{mssb}
\Sigma^{(1b)}(p)=(\frac{\alpha}{4\pi})^2\int_0^1dx\int_0^1dz
\int_0^1du\int_0^1dt\frac{12(p^2-m^2)}{z(1-xz)}
\eeq
\[
\{\frac{1}{\Omega}[\hat p\eta t(1-2x)+\hat px-m(2-x)]
+\hat p(1-x)[\frac{1-t}{\Omega}+t(1-t)\frac{p^2\eta^2}{\Omega^2}]
\]
\[
+\frac{x(1-x)u}{1-xz}\{\frac{\hat
p(2-z)t}{\Omega}-\frac{p^2(1-t)}{\Omega^2}[\frac{\hat p-m}{z}
-2\hat p\eta t]\}\}.
\]

Next we integrate by parts to get rid of denominators $\Omega^2$ (compare
previous section) and obtain

\beq              \label{mssbf}
\Sigma^{(1b)}(p)=(\frac{\alpha}{4\pi})^2\int_0^1dx\int_0^1dz
\int_0^1du\int_0^1dt\{\frac{12(p^2-m^2)}{m^2(\omega+\rho\xi)}
\{\hat p[x+(1-x)(2-3t)
\eeq
\[
+\eta t(1-2x)]-m(2-x)+\frac{x(1-x)u}{1-xz}[\hat
p(2-z)t+2\hat p\frac{1-2t}{\eta}]\}
\]
\[
-(\hat p-m)^2\frac{12p^2(\hat
p+m)x(1-x)u(1-t)}{m^4(\omega+\rho\xi)^2}\}.
\]

One term with denominator $\Omega^2\sim(\omega+\rho\xi)^2$ survived
integration by parts but it already contains explicit factor $(\hat p-m)^2$
and is completely safe under subtraction.

\subsection{Contribution to the Self-Energy Operator Induced by the Term
$R^{(1c)}_\beta$}

Unlike previous terms the term $R^{(1c)}_\beta$ is ultravioletly
divergent and leads to a logarithmically divergent contribution to
the self-energy operator

\beq
\Sigma^{(1c)}(p)=2(\frac{\alpha}{4\pi})^2
\int_0^1dx\int_0^1dz\int\frac{d^4q}{i\pi^2}[\frac{x}{\Delta}C^{(1)}_\beta
+\frac{x^2}{\Delta}C^{(2)}_\beta]\frac{\hat p+\hat q+m}{q^2D(p+q)}
(\gamma^\beta+\frac{2\hat qq^\beta}{q^2}).
\eeq

Since all divergent contributions contain extra powers of integration
momentum $q$ in the numerator we transform numerators prior to integration
separating powers of $q^2$ explicitly. Then we cancel this factor $q^2$ over
similar denominator factor reducing thus the number of Feynman parameters in
the divergent terms. In the end of calculations we will insert additional
integration parameter in these terms in order to obtain integral
representation for the divergent term which does not contain logs. As a
result we will obtain integral representation for the contribution of the
divergent terms to the self-energy operator which contains only four
integration variables and contains denominators which are not too singular
on the mass-shell .

Extraction of the explicit factors $q^2$ is performed with the help of the
identities

\beq
C^{(1)}_\beta(\hat p+m)(\gamma^\beta+\frac{2\hat qq^\beta}{q^2})
=C^{(1)}_\beta(\hat p+m)\gamma^\beta+3\hat p(-3p^2+4m\hat p+2m^2)
\eeq
\[
+6z[m(\hat q\hat p-\hat p\hat q)+3m(\hat p+m)\hat q-2p^2\hat q]
\]
\[
+(3p^2+2m\hat p+4m^2)\frac{2\hat q\hat p\hat q+\hat
pq^2}{q^2}+12z\frac{(pq)\hat q\hat p\hat q}{q^2},
\]
\[
C^{(1)}_\beta\hat q(\gamma^\beta+\frac{2\hat qq^\beta}{q^2})
=q^23z(8m-5\hat p)+2(4m-3\hat p)(\hat p\hat q+2\hat q\hat
p)-z(2\hat q\hat p\hat q+\hat pq^2),
\]
\[
C^{(2)}_\beta(\hat p+m)(\gamma^\beta+\frac{2\hat qq^\beta}{q^2})
=C^{(2)}_\beta(\hat p+m)\gamma^\beta+q^23z^2(2m-\hat p)+3\hat pp^2
\]
\[
+2z(4p^2\hat q+5m\hat p\hat q+m\hat q\hat
p)+3z^2(2\hat q\hat p\hat q+\hat pq^2)
+\hat p(\hat p+2m)\frac{2\hat q\hat p\hat q+\hat pq^2}{q^2}+4z\frac{(pq)\hat
q\hat p\hat q}{q^2},
\]
\[
C^{(2)}_\beta\hat q(\gamma^\beta+\frac{2\hat qq^\beta}{q^2})
=q^2(12z^2\hat q+18z\hat p)+2z(2\hat q\hat p\hat q+q^2\hat p)+4(2\hat p\hat
q\hat p+p^2\hat q).
\]

Then we obtain contribution to the self-energy operator in the form

\beq                \label{massc}
\Sigma^{(1c)}(p)=2(\frac{\alpha}{4\pi})^2\int_0^1dx\int_0^1dz
\int\frac{d^4q}{i\pi^2}\sum_{n=0}^2N_c(1,1,n)G(1,1,n),
\eeq

where

\beq
N_c(1,1,0)=3xz[4\hat qxz+\hat p(-5+6x-xz)+2m(4+xz)],
\eeq
\[
N_c(1,1,1)=[xC^{(1)}_\beta+x^2C^{(2)}_\beta](\hat p+m)\gamma^\beta+3\hat px
(-3p^2+4m\hat p+2m^2)
\]
\[
+2x(4m-3\hat p)(\hat p\hat q+2\hat q\hat p)+3\hat pp^2x^2
+4x^2(2\hat p\hat q\hat p+p^2\hat q)+6xz[m(\hat q\hat p-\hat p\hat q)
\]
\[
+3m(\hat p+m)\hat q-2p^2\hat q]+2x^2z(4p^2\hat q+5m\hat p\hat q+m\hat q\hat p)
+xz(-1+2x+3xz)(2\hat q\hat p\hat q+\hat pq^2),
\]
\[
N_c(1,1,2)=[p^2x(3+x)+2m\hat px(1+x)+4m^2x](2\hat q\hat p\hat q+\hat pq^2)
+4xz(3+x)(pq)\hat q\hat p\hat q.
\]

Performing next momentum integration in \eq{massc} (note that in the term
with numerator $N_c(1,1,0)$ integration momentum is shifted as
$q\rightarrow q-p\eta$ unlike the standard shift $q\rightarrow q-p\eta t$
in all other terms) we obtain

\beq
\Sigma^{(1c)}(p)=-2(\frac{\alpha}{4\pi})^2\int_0^1dx\int_0^1dz
\int_0^1du\int_0^1dt\frac{1}{z(1-xz)}
\eeq
\[
\{{\cal N}_c(1,1,0)(\log{\frac{\Lambda^2}{\Omega_{|t=1}}}-1)+6\hat px^2z^2\eta
-\frac{{\cal N}_c(1,1,1)}{\Omega}
+\frac{(1-t){\cal N}_c(1,1,2)}{t\Omega^2}\},
\]

where

\beq
{\cal N}_c(1,1,0)=3xz[\hat p(-5+6x-xz-4xz\eta)+2m(4+xz)],
\eeq
\[
{\cal N}_c(1,1,1)=\hat pp^2[9x(-3+x)+6x(3-2x)\eta t+10xz(3-2x)\eta t
\]
\[
+3xz(-1+2x+5xz)\eta^2t^2]+mp^26x[6-(4+7z+2xz)\eta t]+m^2\hat p6x(3-5z\eta t),
\]
\[
{\cal N}_c(1,1,2)=\hat pp^2\eta^2t^2[p^2x(3+x)(3-4z\eta t)+6m\hat
px(1+x)+12m^2x]
\]

and

\beq
\Omega_{|t=1}=\Omega(t=1).
\eeq

Once again one may easily get rid of denominator $\Omega^2$ with the help of
integration by parts over $t$. We then obtain

\beq           \label{mass1c}
\Sigma^{(1c)}(p)=-(\frac{\alpha}{4\pi})^22\int_0^1dx\int_0^1dz
\int_0^1du\int_0^1dt\{\frac{{\cal N}_c(1,1,0)}{z(1-xz)}
(\log{\frac{\Lambda^2}{\Omega_{|t=1}}}-1)
\eeq
\[
+\frac{6\hat px^2z^2\eta}{z(1-xz)}
+\frac{{\cal N}'_c(1,1,2)-{\cal N}_c(1,1,1)}{m^2(\omega+\rho\xi)}\},
\]

where

\beq
{\cal N}'_c(1,1,2)=3\hat p(1-2t)x[p^2(3+x)+2m\hat p(1+x)+4m^2]-4\hat
pp^2(2-3t)x(3+x)z\eta t.
\eeq

\section {Calculation of the Contribution to the Mass Operator Induced
by the Term $R^{(2)}_\beta$}

Contribution to the mass operator connected with the term
$R^{(2)}_\beta$ in eq.(\ref{repr}) is the simplest one to calculate. Let
us start with convergent integration over $l$

\beq      \label{r2bet}
R^{(2)}_\beta=-2\int\frac{d^4l}{i\pi^2}\frac{1}{l^2D^2(p_0+l)}
\{4\gamma_\beta[m^2+2\frac{(p_0l)^2}{l^2}]
\eeq
\[
+[4ml_\beta+8(p_0l)\gamma_\beta]-[2\hat
l\gamma_\beta\hat l+l^2\gamma_\beta]\}.
\]

We combine denominators in the same way as in \eq{combg} ($z=0$ now since
there are only two different factors in the denominator)

\beq       \label{combinf}
(1-x)l^2+xD(p_0+l)=(l+xp_0)^2-x\Delta_0,
\eeq

where $\Delta_0=\Delta(z=0)$.

Integrand on the right hand side in \eq{r2bet} is a sum of terms containing
increasing powers of integration momentum $l$ (or powers of the Feynman
parameter $x$ after integration over momentum). Here we encounter the
problem concerning the term with the lowest power of integration
momentum. It is easy to see that if one puts momentum $p_0$ to be exactly on
the mass-shell then respective integral turns out to be the sum of two
infrared divergent terms. These divergences, of course, cancel leading to
a finite result, however, one has to be extremely careful performing this
calculation. Correct way to perform integration here is to preserve
small nonzero virtuality $\rho_0=(m^2-p_0^2)/m^2$ connected with momentum
$p_0$ in the denominator $\Delta_0=m^2[x+\rho_0(1-x)]$ at intermediate
stages of integration. We thus regularize would be infrared divergences,
perform then necessary integrations and go on the mass-shell only in the end
of all calculations

\beq                                 \label{xint}
R^{(2)}_\beta=-2m^2\gamma_\beta
\int_0^1dx\{-4[\frac{2-x}{\Delta_0}-2m^2\frac{x(1-x)}{\Delta_0^2}]
+12\frac{x}{\Delta_0}+3\frac{x^2}{\Delta_0}\}
\eeq
\[
=-3\gamma_\beta.
\]

This simple integration nicely illustrates the necessity to be extremely
careful about going to the mass-shell limit in the presence of the would
be infrared divergences.  Really, if we naively put $p_0=(m,{\bf 0})$
prior to integration over $x$ in eq.(\ref{xint}) we would obtain a finite
result (potentially infrared divergent terms cancel in the integrand even
in this case) but continuity in the external momentum would be lost and,
hence, that result would be wrong.

It is a simple exercise now to obtain respective contribution to the mass
operator

\beq         \label{ured}
\Sigma^{(2)}(p)=-3(\frac{\alpha}{4\pi})^2\int\frac{d^4q}{i\pi^2}
\gamma_\beta\frac{\hat p+\hat q+m}{q^2D(p+q)}(\gamma^\beta+2\frac{\hat
qq^\beta}{q^2}).
\eeq

It is easy to see that integral in eq.(\ref{ured}) only by numerical factor
differs from the expression for the unrenormalized one-loop self-energy
operator in the FY gauge (see, e.g \cite{ann1}). Hence, we may
use results of one-loop calculations and immediately put down
contribution to the renormalized self-energy operator induced by the term
under consideration

\beq    \label{redfin}
\Sigma^{(2)}_R(p)=(\frac{\alpha}{4\pi})^2(\hat p-m)^2\frac{9\hat
p}{m^2}\int_0^1dx\frac{x}{x+\rho(1-x)},
\eeq

and we remind that virtuality $\rho$ was defined in eq.(\ref{auxil}).

\section {Calculation of the Contribution to the Mass Operator Induced
by the Term $R^{(3)}_\beta$}

Consider contribution to the mass operator induced by the term
$R^{(3)}_\beta$ in eq.(\ref{repr}). Integration over $l$ is again
convergent and one easily obtains

\beq
R^{(3)}_\beta=
6\gamma_\beta\int\frac{d^4l}{i\pi^2}[\frac{1}{D(p+l)D(p+l+q)}
-\frac{1}{D^2(p_0+l)}]
\eeq
\[
=6\gamma_\beta\int_0^1dz\log{\frac{m^2}{\Delta_1}}
=-6\gamma_\beta\int_0^1dz\frac{z(1-2z)}{\Delta_1},
\]

where

\beq         \label{delt1}
\Delta_1=\Delta(x=1)\equiv m^2-q^2z(1-z).
\eeq

Respective contribution to the self-energy operator has now the form

\beq
\Sigma^{(3)}(p)=
-6(\frac{\alpha}{4\pi})^2\int_0^1dzz(1-2z)\int\frac{d^4q}{i\pi^2}
\frac{\gamma_\beta(\hat p+\hat q+m)}{\Delta_1D(p+q)}(\gamma^\beta
+2\frac{\hat qq^\beta}{q^2})
\eeq
\[
=-6(\frac{\alpha}{4\pi})^2\int_0^1dzz(1-2z)\int\frac{d^4q}{i\pi^2}
\{3(2m-\hat p)G_1(1,1,0)
\]
\[
+(2\hat q\hat p\hat q+\hat pq^2)G_1(1,1,1)\},
\]

where subscript accompanying function $G_1(1,1,0)$ means that one has to
substitute in the respective definition in \eq{genr}  $\Delta_1$ from
\eq{delt1}.

One may easily perform momentum integration now and obtain

\beq         \label{sum3i}
\Sigma^{(3)}(p)=18(\frac{\alpha}{4\pi})^2\int_0^1dz\frac{1-2z}{1-z}
\int_0^1du\int_0^1dt\{(2m-\hat p)(\log{\frac{\Lambda^2+\Omega_{11}}
{\Omega_{11}}}
\eeq
\[
-\frac{\Lambda^2}{\Lambda^2+\Omega_{11}})-\frac{\hat
pp^2(1-u)^2t^2}{\Omega_{|x=1}}\},
\]

where

\beq
\Omega_{|x=1}=\Omega(x=1)\equiv p^2(1-u)^2t+(m^2-p^2)(1-u)
+\frac{m^2u}{z(1-z)},
\eeq
\[
\Omega_{11}=\Omega(x=1,t=1)\equiv p^2(1-u)^2+(m^2-p^2)(1-u)
+\frac{m^2u}{z(1-z)}.
\]

Note that one has to preserve function $\Omega_{11}$ even on the background
of the infinite cutoff $\Lambda^2$ to facilitate convergence of integration
over Feynman parameters in \eq{sum3i}\footnote{Compare detailed discussion on
this point in \cite{ekse1}.}.

\section{Total Expression for the Overlapping Diagram Contribution to the
Self-Energy Operator}       \label{total}

\subsection{General Representation of the Overlapping Diagram Contribution
to the Self-Energy Operator}

Next task is to perform double subtraction in all contributions to the
self-energy operator obtained above in \eq{mass1a}, \eq{mssbf}, \eq{mass1c}
and \eq{sum3i}. Contribution in \eq{redfin} is already subtracted and does
not need any further transformations. The term containing factor $(\hat
p-m)^2$ in \eq{mssbf} is also already subtracted. All other contributions to
the self-energy operator either contain in the integrand denominator
$\omega+\rho\xi$ or logarithm of the cutoff momentum. Consider first
logarithmically divergent contributions.

Divergent contribution in \eq{mass1c} has the form

\beq           \label{div1c}
\Sigma^{(1c)}_{div}(p)=-(\frac{\alpha}{4\pi})^2\int_0^1dx\int_0^1dz
\int_0^1du\frac{6x}{1-xz}[(\hat p-m)h_1+mh_2]
(\log{\frac{\Lambda^2}{\Omega_{|t=1}}}-1),
\eeq

where

\beq
h_1=-5+6x-xz-4xz\eta,\qquad h_2=3+6x+xz-4xz\eta.
\eeq

Momentum cutoff disappears after subtraction and emerging logarithm may be
put down with the help of integral representation

\beq
\log{\frac{\omega_{|t=1}+\rho\xi_{|t=1}}{\omega_{|t=1}}}=\rho\xi_{|t=1}
\int_0^1\frac{dv}{\omega_{|t=1}+\rho\xi_{|t=1}v}.
\eeq

This parametric representation of the logarithm gives one a chance to
gain an additional apparent factor $\hat p-m$ in the numerator which is very
convenient for the subtraction procedure.

Subtracted expression in \eq{div1c} acquires then the form

\beq
\Sigma^{(1c)}_{div,R}(p)=-(\frac{\alpha}{4\pi})^2\frac{(\hat
p-m)^2}{m^2}\int_0^1dx\int_0^1dz \int_0^1du\frac{6x\xi_{|t=1}}{1-xz}
\eeq
\[
\int_0^1\frac{dv}{\omega_{|t=1}+\rho\xi_{|t=1}v}[\hat p(h_1
+2h_2\frac{\xi_{|t=1}v}{\omega_{|t=1}})+m(h_1+h_2
+2h_2\frac{\xi_{|t=1}v}{\omega_{|t=1}})].
\]

Consider next divergent contribution in \eq{sum3i}. It looks like exactly as
the previous divergent contribution with $h_1=1$ and $h_2=-1$

\beq         \label{divsum3i}
\Sigma^{(3)}_{div}(p)=-18(\frac{\alpha}{4\pi})^2\int_0^1dz\frac{1-2z}{1-z}
\int_0^1du[(\hat p-m)-m]
\eeq
\[
(\log{\frac{\Lambda^2+\Omega_{11}}{\Omega_{11}}}
-\frac{\Lambda^2}{\Lambda^2+\Omega_{11}}).
\]

One easily obtains doubly subtracted contribution to the mass operator

\beq
\Sigma^{(3)}_{div,R}(p)=-18(\frac{\alpha}{4\pi})^2\frac{(\hat
p-m)^2}{m^2}\int_0^1dz\frac{1-2z}{1-z}
\int_0^1du\xi_{11}\int_0^1\frac{dv}{\omega_{11}+\rho\xi_{11}v}
\eeq
\[
[\hat p(1-2\frac{\xi_{11}v}{\omega_{11}})-2m\frac{\xi_{11}v}{\omega_{11}}],
\]

where

\[
\xi_{11}\equiv \xi(x=1,t=1)=z(1-z)u(1-u),
\]
\[
\omega_{11}\equiv \omega(x=1,t=1)=u+z(1-z)(1-u)^2.
\]

Now we have to consider subtraction of all other entries in \eq{mass1a},
\eq{mssbf}, \eq{mass1c} and \eq{sum3i} which contain factor
$\omega+\rho\xi$ in the denominator. Collecting all such terms we obtain the
expression

\beq
\Sigma_{|\omega+\rho\xi}(p)=(\frac{\alpha}{4\pi})^2\int_0^1dx\int_0^1dz
\int_0^1du\int_0^1dt\frac{12}{\omega+\rho\xi}\{\frac{(\hat p-m)^2}{m^2}[\hat
pf_{1p}+mf_{1m}]
\eeq
\[
+(\hat p-m)f_2+mf_3\},
\]

where

\beq       \label{auxfun}
f_{i}=f'_{i}+\frac{x(1-x)u}{1-xz}f''_{i},
\eeq
\[
f'_{1p}=2+x(-7+x+3t+xt)+[-2+x(4-2x+9z-2xz-6zt-2xzt)]\eta t+S,
\]
\[
f'_{1m}=2+x(-8+x+5t+4xt)+[-1+x(4-4x+11z-6xz-12zt-4xzt)]\eta t+2S,
\]
\[
f'_2=x(-7+x+8t+7xt)+[-1+x(6-6x+8z-10xz-18zt-6xzt)]\eta t+3S,
\]
\[
f'_3=3x(1+x)t+x(2-2x-3z-4xz-6zt-2xzt)\eta t+S,
\]
\[
f''_{1p}=t(1-6z+10zt-2z\eta t)+4\frac{1-2t}{\eta},
\]
\[
f''_{1m}=t(2-11z+20zt-4z\eta t)+6\frac{1-2t}{\eta},
\]
\[
f''_2=t(1-15z+30zt-6z\eta t)+6\frac{1-2t}{\eta},
\]
\[
f''_3=-t(1+4z-10zt+2z\eta t),
\]
\[
S=\frac{1}{2}xz(-1+2x+5xz)\eta^2t^2.
\]

After subtractions we obtain

\beq
\Sigma_{|\omega+\rho\xi,R}(p)=(\frac{\alpha}{4\pi})^2\frac{(\hat p-m)^2}{m^2}
\int_0^1dx\int_0^1dz
\int_0^1du\int_0^1dt\frac{1}{\omega+\rho\xi}\{\hat
p[f_{1p}+f_2\frac{\xi}{\omega}
\eeq
\[
+2f_3(\frac{\xi}{\omega})^2]
+m[f_{1m}+(f_2+f_3)\frac{\xi}{\omega}+2f_3(\frac{\xi}{\omega})^2]\}.
\]

Subtraction in the last term in \eq{sum3i} may be easily performed finishing
thus calculation of all contributions to the subtracted self-energy
operator induced by the diagram with overlapping photons.

Collecting all entries to the self-energy we obtain

\beq      \label{totover}
\Sigma_R(p)=(\frac{\alpha}{4\pi})^2\frac{(\hat
p-m)^2}{m^2}\int_0^1dx\int_0^1dz\int_0^1du\int_0^1dt\int_0^1dv
\{\frac{12}{\omega+\rho\xi}\{\hat p[f_{1p}+f_2\frac{\xi}{\omega}
\eeq
\[
+2f_3(\frac{\xi}{\omega})^2]
+m[f_{1m}+(f_2+f_3)\frac{\xi}{\omega}+2f_3(\frac{\xi}{\omega})^2]
-\frac{p^2(\hat p+m)x(1-x)u(1-t)}{m^2(\omega+\rho\xi)}\}
\]
\[
-\frac{6x}{1-xz}\frac{\xi_{|t=1}}{\omega_{|t=1}+\rho\xi_{|t=1}v}
[\hat p(h_1+2h_2\frac{\xi_{|t=1}v}{\omega_{|t=1}})+m(h_1+h_2
+2h_2\frac{\xi_{|t=1}v}{\omega_{|t=1}})]
\]
\[
-18\frac{z(1-2z)(1-u)^2t^2}{\omega_{|x=1}+\rho\xi_{|x=1}}
\{\hat p[1+3\frac{\xi_{|x=1}}{\omega_{|x=1}}
+2(\frac{\xi_{|x=1}}{\omega_{|x=1}})^2]
+2m(1+\frac{\xi_{|x=1}}{\omega_{|x=1}})^2\}
\]
\[
-18\frac{1-2z}{1-z}\frac{\xi_{11}}{\omega_{11}+\rho\xi_{11}v}
[\hat p(1-2\frac{\xi_{11}v}{\omega_{11}})-2m\frac{\xi_{11}v}{\omega_{11}}]
+9\frac{\hat px}{x+\rho(1-x)}\}
\]
\[
\equiv (\frac{\alpha}{4\pi})^2\frac{(\hat
p-m)^2}{m^2}[\hat p\sigma_p(\rho)+m\sigma_m(\rho)].
\]

Note that despite the appearance each term on the right hand side is at most
a four dimensional integral.

\subsection{Infrared Safe Representation of the Overlapping Diagram
Contribution to the Self-Energy Operator}

The representation for the overlapping diagram contribution in \eq{totover}
is quite suitable for calculation of the respective contribution to HFS but
needs further transformations for the Lamb shift calculation. The problem
emerges when one tries to calculate values of the subtraction constants
$\sigma_p(0)$ and $\sigma_m(0)$ which enter the expression for the Lamb
shift (see below). Integrals containing terms $f_2\xi/\omega$ and
$2f_3\xi^2/\omega^2$ in \eq{totover} diverge as $\log\rho$ when $\rho$ goes
to zero (we remind that according to \eq{auxil}
$\omega=xu+zt(1-xz)+O(xuzt)$) and this divergence disappears only in the sum

\beq         \label{redpr}
M(\rho)\equiv \int_0^1dx\int_0^1dz\int_0^1du\int_0^1dt
\frac{1}{\omega+\rho\xi}[\frac{f_2\xi}{\omega}
+2(\frac{f_3\xi}{\omega})^2]
\eeq

which enters \eq{totover}.

Throughout investigation of the integrand in \eq{redpr} with the help of the
definitions in \eq{auxil} and \eq{auxfun}  shows that only terms

\beq      \label{infrfun}
(f_2\xi)_{IR}=x(-7+x)z(1-xz)-tu,
\eeq
\[
(f_3\xi^2)_{IR}=xt[(5+x-3z-4xz)z^2(1-xz)^2+u^2(5-u)]
\]

lead to divergencies at vanishing virtuality. Each term in the subtracted
numerator $f_2\xi-(f_2\xi)_{IR}$ contains either factor $xu$ or $zt$ which
suppress divergency at vanishing virtuality. In the same way each term in
the other subtracted numerator $f_3\xi^2-(f_3\xi^2)_{IR}$ contains one of
the factors $(xu)^2$, $(zt)^2$ or $(xu)(zt)$ and respective integrals are
also infrared safe at vanishing virtuality.

We are going to transform integral representation in \eq{totover} to a form
containing integrands which lead only to the integrals finite at $\rho=0$.
Cancellation of infrared divergent terms may be greatly facilitated by
reducing the multiplicity of integration. To reduce the number of
integrations we note that infrared properties of the integrands
does not change if one substitutes in the numerators in \eq{infrfun}
$x\rightarrow \partial\omega/\partial u$ and $tu\rightarrow
u\partial\omega/\partial z$. Using this fact we separate infrared divergent
terms by substituting these derivatives in \eq{infrfun} and define new
infrared safe functions

\beq
\widetilde{f_2\xi}\equiv f_2\xi-{\cal D}_2\omega,
\eeq
\[
\widetilde{f_3\xi^2}\equiv f_3\xi^2-{\cal D}_3\omega,
\]

where

\beq
{\cal D}_2=(-7+x)z(1-xz)\frac{\partial}{\partial
u}-u\frac{\partial}{\partial z},
\eeq
\[
{\cal D}_3=t(5+x-3z-4xz)z^2(1-xz)^2\frac{\partial}{\partial
u}+xu^2(5-u)\frac{\partial}{\partial z}.
\]

Now $M(\rho)$  in \eq{redpr} acquires the form

\beq        \label{tildeex}
M(\rho)\equiv \widetilde{M(\rho)}+M_1(\rho),
\eeq

where $\widetilde{M(\rho)}$ differs from $M(\rho)$ only due to substitutions
$f_2\xi\rightarrow\widetilde{f_2\xi}$ and
$f_3\xi^2\rightarrow\widetilde{f_3\xi^2}$.  Term $\widetilde{M(\rho)}$ is
infrared safe and logarithmic infrared divergency is connected only with the
term

\beq
M_1(\rho)=\int_0^1dx\int_0^1dz\int_0^1du\int_0^1dt
\frac{1}{\omega+\rho\xi}[\frac{{\cal D}_2\omega}{\omega}
+\frac{2{\cal D}_3\omega}{\omega^2}]
\eeq
\[
=\int_0^1dx\int_0^1dz\int_0^1du\int_0^1dt\int_0^1dv
[\frac{{\cal D}_2\omega}{(\omega+\rho v\xi)^2}
+2(1-v)\frac{2{\cal D}_3\omega}{(\omega+\rho v\xi)^3}].
\]

We easily extract infrared safe part $M_\xi(\rho)$ from the term $M_1(\rho)$
with the help of the trivial substitution

\beq     \label{trivid}
\omega=-\rho v\xi+(\omega+\rho v\xi),
\eeq

which is valid according to the definition in \eq{auxil}.

First term on the right hand side in \eq{trivid} produces infrared safe
contribution to  $M_1(\rho)$

\beq
\rho M_\xi(\rho)=-\rho\int_0^1dx\int_0^1dz\int_0^1du\int_0^1dt\int_0^1dvv
[\frac{{\cal D}_2\xi}{(\omega+\rho v\xi)^2}
+4(1-v)\frac{{\cal D}_3\xi}{(\omega+\rho v\xi)^3}],
\eeq

while the second term on the right hand side contains complete derivatives
over one of the Feynman parameters which may be easily integrated. Hence, we
obtain the following representation for the $M_1(\rho)$ term

\beq
M_1(\rho)=\rho M_\xi(\rho)+M_{up}(\rho)+M_{down}(\rho),
\eeq

where

\beq
M_{up}(\rho)=\int_0^1dx\int_0^1dz\int_0^1du\int_0^1dt\int_0^1dv
\{\frac{(7-x)z(1-xz)}{\omega_{|u=1}+\rho v\xi_{|u=1}}
\eeq
\[
+\frac{u}{\omega_{|z=1}+\rho v\xi_{|z=1}}
-2(1-v)[\frac{t(5+x-3z-4xz)z^2(1-xz)^2}{(\omega_{|u=1}+\rho v\xi_{|u=1})^2}
\]
\[
+\frac{xu^2(5-u)}{(\omega_{|z=1}+\rho v\xi_{|z=1})^2}]\},
\]
\[
M_{down}(\rho)= \int_0^1dx\int_0^1dz\int_0^1du\int_0^1dt\int_0^1dv
\{\frac{-7+x}{t+\rho v(1-t)}
\]
\[
-\frac{1}{x+\rho v(1-x)}
+2(1-v)[\frac{t(5+x-3z-4xz)}{[t+\rho v(1-t)]^2}
+\frac{x(5-u)}{[x+\rho v(1-x)]^2}]\}.
\]

It is easy to check that all integrands in $M_{up}(\rho)$ are infrared safe
even at vanishing virtuality, while each integrand in the
expression for the term $M_{down}(\rho)$ produces $\log\rho$ in the small
$\rho$ limit. Happily due to simplicity of parametric integrals in the
expression for $M_{down}(\rho)$  one may easily obtain its parametric
representation in the form of one-dimensional integral

\beq
M_{down}(\rho)=\frac{15}{2}\int_0^1dv\{-\frac{1}{2}-\frac{\rho v}{1-\rho v}
\ln\frac{1}{\rho v}
\eeq
\[
+2(1-v)[\frac{\rho v}{1-\rho v}\ln\frac{1}{\rho v}
+\frac{\rho v}{1-\rho v}(\frac{1}{1-\rho v}\ln\frac{1}{\rho v}-1)]\}.
\]

It is easy to see that total expression for $M_{down}(\rho)$ at
$\rho\rightarrow 0$ is free of logs and one may easily obtain

\beq
M_{down}(0)=-\frac{15}{4}.
\eeq

Thus, we have obtained representation for the contribution of the
overlapping diagram to the self-energy operator in the form

\beq           \label{safered}
\Sigma_R(p)=(\frac{\alpha}{4\pi})^2\frac{(\hat
p-m)^2}{m^2}\{\hat p\widetilde{\sigma}_p(\rho)+m\widetilde{\sigma}_m(\rho)
\eeq
\[
+12(\hat p+m)[\rho M_\xi(\rho)+M_{up}(\rho)+M_{down}(\rho)]\},
\]

where $\widetilde{\sigma}_p(\rho)$ and $\widetilde{\sigma}_m(\rho)$ differ
from the respective expressions in \eq{totover}  only by substitutions
$f_2\xi\rightarrow\widetilde{f_2\xi}$ and
$f_3\xi^2\rightarrow\widetilde{f_3\xi^2}$ (compare \eq{tildeex}).

Each term on the right hand side in \eq{safered} is explicitly finite at
vanishing virtuality. Hence, this representation for the self-energy
operator is very convenient for subtraction of its value on the mass shell
and for calculation of the respective contribution to the Lamb shift.

\section{Contributions to the Energy Splittings induced by the
Overlapping Mass Operator}
\subsection{Contribution to HFS Splitting}

Radiative correction to hyperfine splitting induced by the graph in
Fig.1 is given by the matrix element of this diagram calculated with on
mass-shell external electron lines and projected on respective spin states
and multiplied by the square in the origin of the Schr\"{o}dinger-Coulomb
wave function (for more details see, e.g. \cite{ekse1} and \cite{ann1}). It
is not difficult to obtain explicit expression to the energy splitting
having formula for the mass operator in
eq.(\ref{totover})\footnote{Integration momentum $k$ is measured in the
units of electron mass in this section.}

\beq
\Delta E_{HFS}=\frac{\alpha^2(Z\alpha)}{\pi
n^3}E_F(-\frac{1}{2\pi^2})\int_0^\infty d|{\bf k}|\:\sigma_p({\bf k}^2)
\eeq
\[
=\frac{\alpha^2(Z\alpha)}{\pi
n^3}E_F(-\frac{1}{2\pi^2})\{\frac{9\pi^2}{4}+\pi {\cal H}\},
\]

where

\beq
{\cal H}=\int_0^1dx\int_0^1dz\int_0^1du\int_0^1dt\{
\frac{6}{\sqrt{\omega\xi}}[f_{1p}+f_2\frac{\xi}{\omega}
\eeq
\[
+2f_3(\frac{\xi}{\omega})^2]
+\frac{3x(1-x)u(1-t)}{\sqrt{\omega\xi}}(\frac{1}{\xi}-\frac{1}{\omega})
\]
\[
-\frac{2x}{1-xz}\sqrt{\frac{\xi_{|t=1}}{\omega_{|t=1}}}
(3h_1+2h_2\frac{\xi_{|t=1}}{\omega_{|t=1}})
\]
\[
-9\frac{z(1-2z)(1-u)^2t^2}{\sqrt{\omega_{|x=1}\xi_{|x=1}}}
[1+3\frac{\xi_{|x=1}}{\omega_{|x=1}}
+2(\frac{\xi_{|x=1}}{\omega_{|x=1}})^2]
\]
\[
-6\frac{1-2z}{1-z}\sqrt{\frac{\xi_{11}}{\omega_{11}}}
(3-2\frac{\xi_{11}}{\omega_{11}})\}.
\]

Numerically we obtain

\beq
\Delta E_{HFS}=-1.984(1)\frac{\alpha^2(Z\alpha)}{\pi n^3}E_F.
\eeq

\subsection{Contribution to the Lamb Shift}

Contribution to the Lamb shift induced by the diagram in Fig.1 is given by
the matrix element similar to the one in the case of HFS, the only
difference being in the spin structure and the necessity to perform an
additional subtraction on the mass shell, already mentioned above (see, e.g.
\cite{ekse1} and \cite{ego}). This subtraction is greatly facilitated by the
representation in eq.(\ref{safered}).
Using this representation we obtain

\beq     \label{lambk}
\Delta E_L=\frac{8(Z\alpha)^5}{\pi
n^3}(\frac{\alpha}{4\pi})^2m(\frac{m_r}{m})^3 \int_0^\infty
\frac{d|{\bf k}|}{{\bf k}^2}
\{\{[\widetilde{\sigma}_p({\bf k}^2)+\widetilde{\sigma}_m({\bf
k}^2)]-[\widetilde{\sigma}_p(0)+\widetilde{\sigma}_m(0)]\}
\eeq
\[
+24{\bf k}^2M_\xi({\bf k}^2)+24[M_{up}({\bf k}^2)-M_{up}(0)]
+24[M_{down}({\bf k}^2)-M_{down}(0)]\}
\]
\[
\equiv \frac{\alpha^2(Z\alpha)^5}{\pi n^3}m(\frac{m_r}{m})^3\frac{1}{2\pi^2}
\int_0^\infty{d|{\bf k}|}[\Delta{\cal E}_{L1}({\bf k}^2)
+\Delta{\cal E}_{L2}({\bf k}^2)+\Delta{\cal E}_{L3}({\bf k}^2)
+\Delta{\cal E}_{L4}({\bf k}^2)].
\]

According to definitions in \eq{totover} and \eq{safered}

\beq
\widetilde{\sigma}_p({\bf k}^2)+\widetilde{\sigma}_m({\bf
k}^2)=\int_0^1dx\int_0^1dz\int_0^1du
\int_0^1dt\int_0^1dv\{\frac{12}{\omega+{\bf k}^2\xi}
\eeq
\[
[f_{1p}+f_{1m}+f_3\frac{\xi}{\omega}+2(\frac{\widetilde{f_2\xi}}{\omega}
+\frac{\widetilde{2f_3\xi^2}}{\omega^2})]
-\frac{24(1-{\bf k}^2)x(1-x)u(1-t)}{(\omega+{\bf k}^2\xi)^2}
\]
\[
-\frac{6x}{1-xz}\frac{\xi_{|t=1}}{\omega_{|t=1}+{\bf k}^2\xi_{|t=1}v}
(2h_1+h_2+4h_2\frac{\xi_{|t=1}v}{\omega_{|t=1}})
\]
\[
-18\frac{z(1-2z)(1-u)^2t^2}{\omega_{|x=1}+{\bf k}^2\xi_{|x=1}}
[3+7\frac{\xi_{|x=1}}{\omega_{|x=1}}
+4(\frac{\xi_{|x=1}}{\omega_{|x=1}})^2]
\]
\[
-18\frac{1-2z}{1-z}\frac{\xi_{11}}{\omega_{11}+{\bf k}^2\xi_{11}v}
(1-4\frac{\xi_{11}v}{\omega_{11}})
+9\frac{x}{x+{\bf k}^2(1-x)}\}.
\]

Subtraction may be easily performed with the help of identities

\beq
\frac{1}{{\bf k}^2}[\frac{1}{\omega+{\bf k}^2\xi}-\frac{1}{\omega}]
=-\frac{\xi}{\omega}\frac{1}{\omega+{\bf k}^2\xi},
\eeq
\[
\frac{1}{{\bf k}^2}[\frac{1}{(\omega+{\bf k}^2\xi)^2}-\frac{1}{\omega^2}]
=-\frac{\xi}{\omega}[\frac{1}{(\omega+{\bf k}^2\xi)^2}
+\frac{1}{\omega(\omega+{\bf k}^2\xi)}],
\]

and we obtain

\beq
\Delta{\cal E}_{L1}({\bf k}^2)
=\int_0^1dx\int_0^1dz\int_0^1du\int_0^1dt\int_0^1dv
\eeq
\[
\{-\frac{12}{\omega+{\bf
k}^2\xi}\frac{\xi}{\omega}[f_{1p}+f_{1m}+f_3\frac{\xi}{\omega}
+2(\frac{\widetilde{f_2\xi}}{\omega}+\frac{\widetilde{2f_3\xi^2}}{\omega^2})]
\]
\[
+\frac{24x(1-x)u(1-t)}{(\omega+{\bf k}^2\xi)^2}[1+2\frac{\xi}{\omega}
+{\bf k}^2(\frac{\xi}{\omega})^2]
\]
\[
+\frac{6x}{1-xz}\frac{\xi_{|t=1}}{\omega_{|t=1}+{\bf k}^2\xi_{|t=1}v}
\frac{\xi_{|t=1}v}{\omega_{|t=1}}
(2h_1+h_2+4h_2\frac{\xi_{|t=1}v}{\omega_{|t=1}})
\]
\[
+18\frac{z(1-2z)(1-u)^2t^2}{\omega_{|x=1}+{\bf k}^2\xi_{|x=1}}
\frac{\xi_{|x=1}}{\omega_{|x=1}}[3+7\frac{\xi_{|x=1}}{\omega_{|x=1}}
+4(\frac{\xi_{|x=1}}{\omega_{|x=1}})^2]
\]
\[
+18\frac{1-2z}{1-z}\frac{\xi_{11}}{\omega_{11}+{\bf k}^2\xi_{11}v}
\frac{\xi_{11}v}{\omega_{11}}(1-4\frac{\xi_{11}v}{\omega_{11}})
-9\frac{1-x}{x+{\bf k}^2(1-x)}\}.
\]

Integration over $v$ is greatly facilitated by the obvious relations

\beq             \label{vint}
\int_0^1dv\int_0^\infty\frac{d|{\bf k}|}{{\bf k}^2}F({\bf k}^2v)
=\frac{2}{3}\int_0^\infty\frac{d|{\bf k}|}{{\bf k}^2}F({\bf k}^2),
\eeq
\[
\int_0^1dvv\int_0^\infty\frac{d|{\bf k}|}{{\bf k}^2}F({\bf k}^2v)
=\frac{2}{5}\int_0^\infty\frac{d|{\bf k}|}{{\bf k}^2}F({\bf k}^2),
\]

which are valid for arbitrary function F. After integration over $v$ and
${\bf k}$  we obtain

\beq    \label{1cont}
\int_0^\infty{d|{\bf k}|}
\Delta{\cal E}_{L1}({\bf k}^2)=-\frac{9\pi^2}{4}+\pi L,
\eeq

where

\beq
L=\int_0^1dx\int_0^1dz\int_0^1du\int_0^1dt
\{-\frac{6}{\sqrt{\omega\xi}}\frac{\xi}{\omega}[f_{1p}+f_{1m}
+f_3\frac{\xi}{\omega}
\eeq
\[
+2(\frac{\widetilde{f_2\xi}}{\omega}+\frac{\widetilde{2f_3\xi^2}}{\omega^2})]
+\frac{6x(1-x)u(1-t)}{\omega^\frac{3}{2}\sqrt{\xi}}(1+3\frac{\xi}{\omega})
\]
\[
+\frac{2x}{1-xz}(\frac{\xi_{|t=1}}{\omega_{|t=1}})^\frac{3}{2}
(2h_1+h_2+\frac{12}{5}h_2\frac{\xi_{|t=1}}{\omega_{|t=1}})
\]
\[
+9\frac{z(1-2z)(1-u)^2t^2}{\sqrt{\omega_{|x=1}\xi_{|x=1}}}
\frac{\xi_{|x=1}}{\omega_{|x=1}}[3+7\frac{\xi_{|x=1}}{\omega_{|x=1}}
+4(\frac{\xi_{|x=1}}{\omega_{|x=1}})^2]
\]
\[
+6\frac{1-2z}{1-z}(\frac{\xi_{11}}{\omega_{11}})^\frac{3}{2}
(1-\frac{12}{5}\frac{\xi_{11}}{\omega_{11}})\}.
\]

Next we perform integration over $v$ and $|{\bf k}|$ in the second term
on the right hand side in \eq{lambk}

\beq   \label{2cont}
\int_0^\infty{d|{\bf k}|}\Delta{\cal E}_{L2}({\bf k}^2)
=24\int_0^\infty{d|{\bf k}|}M_\xi({\bf k}^2)
\eeq
\[
=-24\int_0^1dx\int_0^1dz\int_0^1du\int_0^1dt\int_0^1dvv
\int_0^\infty{d|{\bf k}|}
[\frac{{\cal D}_2\xi}{(\omega+{\bf k}^2 v\xi)^2}
\]
\[
+4(1-v)\frac{{\cal D}_3\xi}{(\omega+{\bf k}^2 v\xi)^3}]
=-4\pi\int_0^1dx\int_0^1dz\int_0^1du\int_0^1dt\frac{1}{\sqrt\xi
\omega^\frac{3}{2}}[{\cal D}_2\xi+\frac{6}{5}\frac{{\cal D}_3\xi}{\omega}]
\]
\[
\equiv \pi L_\xi.
\]

Contribution of the third term on the right hand side in \eq{lambk} is equal
to

\beq           \label{3cont}
\int_0^\infty{d|{\bf k}|}\Delta{\cal E}_{L3}({\bf k}^2)
=24\int_0^\infty\frac{{d|{\bf k}|}}{{\bf k}^2}[M_{up}({\bf k}^2)
-M_{up}(0)]
\eeq
\[
=24\int_0^1dx\int_0^1dz\int_0^1du\int_0^1dt\int_0^1dv
\int_0^\infty{d|{\bf k}|}
\{-\frac{v\xi_{|u=1}}{\omega_{|u=1}}
\frac{(7-x)z(1-xz)}{\omega_{|u=1}+{\bf k}^2 v\xi_{|u=1}}
\]
\[
-\frac{v\xi_{|z=1}}{\omega_{|z=1}}
\frac{u}{\omega_{|z=1}+{\bf k}^2 v\xi_{|z=1}}
+2(1-v)[\frac{v\xi_{|u=1}}{\omega_{|u=1}}
[\frac{1}{(\omega_{|u=1}+{\bf k}^2 v\xi_{|u=1})^2}
\]
\[
+\frac{1}{\omega(\omega_{|u=1}+{\bf k}^2 v\xi_{|u=1})}]
t(5+x-3z-4xz)z^2(1-xz)^2
\]
\[
+\frac{v\xi_{|z=1}}{\omega_{|z=1}}
[\frac{1}{(\omega_{|z=1}+{\bf k}^2 v\xi_{|z=1})^2}+
\frac{1}{\omega(\omega_{|z=1}+{\bf k}^2 v\xi_{|z=1})}]
xu^2(5-u)\}
\]
\[
=8\pi\int_0^1dx\int_0^1dz\int_0^1dt\frac{z(1-xz)\sqrt{\xi_{|u=1}}}
{\omega_{|u=1}^\frac{3}{2}}
\{-7+x+\frac{6}{5}\frac{t(5+x-3z-4xz)z(1-xz)}{\omega_{|u=1}}\}
\]
\[
+8\pi\int_0^1dx\int_0^1du\int_0^1dt\frac{u\sqrt{\xi_{|z=1}}}
{\omega_{|z=1}^\frac{3}{2}}\{-1+\frac{6}{5}\frac{xu(5-u)}{\omega_{|z=1}}\}
\]
\[
\equiv \pi L_{up}.
\]

Last term on the right hand side in \eq{lambk} gives

\beq                   \label{4cont}
\int_0^\infty{d|{\bf k}|}\Delta{\cal E}_{L4}({\bf k}^2)
=24\int_0^\infty\frac{{d|{\bf k}|}}{{\bf k}^2}[M_{down}({\bf k}^2)
-M_{down}(0)]
\eeq
\[
=180\int_0^1dvv\int_0^\infty\frac{d|{\bf k}|}{1-{\bf k}^2 v}
\{\ln({\bf k}^2 v)-2(1-v)[\ln({\bf k}^2 v)+1+
+\frac{\ln({\bf k}^2 v)}{1-{\bf k}^2 v}]\}
\]
\[
=24\int_0^\infty{d|{\bf k}|}\{\frac{\ln{\bf k}^2}{1-{\bf k}^2}
-\frac{4}{1-{\bf k}^2}[1+\frac{\ln{\bf k}^2}{1-{\bf k}^2}]\}
\]
\[
=12\pi^2.
\]

Combining expressions in \eq{1cont}, \eq{2cont}, \eq{3cont} and
\eq{4cont} we obtain final expression for the contribution to the Lamb shift
in the form

\beq
\Delta E_L=m(\frac{m_r}{m})^3\frac{\alpha^2(Z\alpha)^5}{\pi n^3}
\frac{1}{2\pi^2}\{\frac{39\pi^2}{4}+\pi(L+L_\xi+L_{up})\},
\eeq

or numerically

\beq
\Delta E_L=1.749(2)\frac{\alpha^2(Z\alpha)^5}{\pi n^3}(\frac{m_r}{m})^3m.
\eeq

\bigskip

This work was supported, in part, by Soros Humanitarian Foundations Grants
awarded by the American Physical Society, and by the grant \#93-02-3853
of the Russian Foundation for Fundamental Research.

\newpage

{\bf Figure Caption}

\bigskip\bigskip

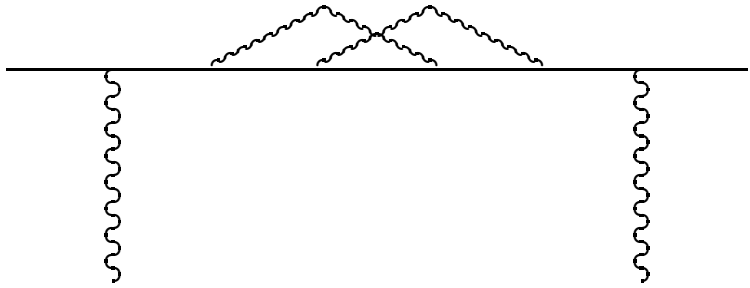
\noindent {\it Fig.1.} Diagram with two external photons and overlapping
two-loop electron self-energy insertion in the electron line.

\newpage
\begin{figure}
 \begin{picture}(100,380)
  \put(80,340){\begin{picture}(200,240)
             \thicklines
             \multiput(80,32.5)(5,2.5){9}{\oval(5,2.5)[lt]}
             \multiput(80,35)(5,2.5){8}{\oval(5,2.5)[br]}
             \multiput(125,52.5)(5,-2.5){8}{\oval(5,2.5)[bl]}
             \multiput(120,52.5)(5,-2.5){9}{\oval(5,2.5)[tr]}
             \multiput(120,32.5)(5,2.5){9}{\oval(5,2.5)[lt]}
             \multiput(120,35)(5,2.5){8}{\oval(5,2.5)[br]}
             \multiput(165,52.5)(5,-2.5){8}{\oval(5,2.5)[bl]}
             \multiput(160,52.5)(5,-2.5){9}{\oval(5,2.5)[tr]}
             \multiput(0,30)(0,90){1}{\line(1,0){280}}
             \multiput(40,27.5)(0,-10){8}{\oval(5,5)[ll]}
             \multiput(40,22.5)(0,-10){8}{\oval(5,5)[rr]}
             \multiput(240,27.5)(0,-10){8}{\oval(5,5)[ll]}
             \multiput(240,22.5)(0,-10){8}{\oval(5,5)[rr]}
             \put(120,-100){Fig.1}
             \end{picture}}
 \end{picture}
\end{figure}

\end{document}